\documentclass[doublecol]{epl2}

\title{Homogeneous {\it vs.} inhomogeneous coexistence of magnetic order and superconductivity probed by NMR in Co and K doped iron pnictides}

\shorttitle{Coexistence of magnetic order and superconductivity in K and Co doped iron pnictides} 

\author{M.-H. Julien\inst{1} \and H.~Mayaffre\inst{1} \and M.~Horvati\'c\inst{2} \and C.~Berthier\inst{2}
\and X. D. Zhang\inst{3} \and W. Wu\inst{3} \and G.F. Chen\inst{3} \and N.L. Wang\inst{3} \and J.L. Luo\inst{3}} \shortauthor{M.-H.~Julien
\etal}
\institute{
  \inst{1} Laboratoire de Spectrom\'etrie Physique, UMR5588 CNRS and Universit\'e J. Fourier - Grenoble, 38402 Saint Martin
d'H\`{e}res, France\\
  \inst{2} Laboratoire National des Champs Magn\'etiques Intenses, UPR 3228 CNRS, 25 avenue des martyrs, BP 166, 38042 Grenoble Cedex 9, France\\
\inst{3} Beijing National Laboratory for Condensed Matter Physics and Institute of Physics, Chinese Academy of Science, P.O. Box 603,
Beijing 100190, China }


\abstract{In Ba(Fe$_{0.95}$Co$_{0.05}$)$_2$As$_2$ all of the $^{75}$As NMR intensity at the paramagnetic resonance position vanishes
abruptly below $T^{\rm onset}_{\rm SDW}=56$~K, indicating that magnetic (spin density wave) order is present in all of the sample volume,
despite bulk superconductivity below $T_c$=15~K. The two phases thus coexist homogeneously at the microscopic scale. In
Ba$_{0.6}$K$_{0.4}$Fe$_2$As$_2$, on the other hand, the signal loss below $T^{\rm onset}_{\rm SDW}\simeq$75~K is not complete, revealing
that magnetic order is bound to finite-size areas of the sample, while the remaining NMR signal shows a clear superconducting response
below $T_c=37$~K. Thus, the two phases are not homogeneously mixed, at least for this potassium concentration. For both samples, spatial
electronic and/or magnetic inhomogeneity is shown to characterize the NMR properties in the normal state.}

\begin{document}

\maketitle

\section{Introduction}

How long range magnetic order can coexist with bulk superconductivity is a central question in a number of unconventional superconductors.
In both families of high temperature superconductors, the copper oxides and the "new" iron pnictides, the superconducting phase is
obtained by adding charge carriers into a phase with antiferromagnetic order. Observation of phase coexistence in such a situation
immediately raises two important questions. The first one is related to the intrinsic character of the coexistence. This question is
obviously linked to a rather subtle issue in these off-stoichiometric materials: to which extent is chemical inhomogeneity of the samples
intrinsic or not ({\it i.e.} unavoidable or not)? The next important question is: Do the phases overlap in space, or do they occupy
mutually exclusive volumes? In short, on which length scale do the two phases coexist?  This is an equally delicate problem because
unambiguous direct experimental proofs are in most cases difficult to obtain, even for true local probes such as nuclear magnetic
resonance (NMR), muon spin rotation ($\mu$SR) or scanning tunneling microscopy (STM). As a matter of fact, the debate is still going on in
the cuprates~(see \cite{Mitrovic08} and references therein). In the iron pnictides, magnetic order (a spin density wave state) has
recently been found to coexist with superconductivity in several (but apparently not all) families, such as Ba$_{1-x}$K$_{x}$Fe$_2$As$_2$
or Ba(Fe$_{1-x}$Co$_{x}$)$_2$As$_2$. Yet, a global picture of the conditions for this coexistence in the pnictides is still lacking.
Because there is an important number of reports on this matter, we primarily refer the reader to the papers presenting phase diagrams of
these systems, see\cite{Zhao08,Luetkens09,Drew09,Ni08,Chen09,Chu09} and references therein.

In this Letter, we investigate single crystals of Ba$_{0.6}$K$_{0.4}$Fe$_2$As$_2$ and Ba(Fe$_{1.95}$Co$_{0.05}$)$_2$As$_2$ with $^{75}$As
nuclear magnetic resonance (NMR). Although both of these samples exhibit coexistence of bulk superconductivity and spin density wave (SDW)
order, the contrasting behaviors observed in their NMR properties shows that the coexistence occurs on different length scales in the two
samples.

\section{Experimental details and bulk properties}

Single crystals Ba$_{0.6}$K$_{0.4}$Fe$_2$As$_2$ and Ba(Fe$_{1.95}$Co$_{0.05}$)$_2$As$_2$ (initial Co content of 0.07) were grown using a
self-flux technique, as detailed elsewhere\cite{Chen08}. The magnetization was measured in a SQUID magnetometer in a 10 Oe magnetic field
applied perpendicular to the $c$-axis after zero-field cooling. The in-plane resistivity was measured using standard four-probe method.
$^{75}$As NMR experiments were performed with a home built pulsed spectrometer.

Figure 1 shows the temperature dependent magnetic susceptibility $\chi$ and the superconducting transitions at $T_c$ for both samples. The
values of onset $T_c$ are $\sim$ 37 K for Ba$_{0.6}$K$_{0.4}$Fe$_2$As$_2$ and 15 K for Ba(Fe$_{1.95}$Co$_{0.05}$)$_2$As$_2$, respectively.
The same values of $T_c$ are obtained from measurements of the in-plane resistivity $\rho_{ab}$ (see figure 3(a)). These values of $T_c$
are in agreement with the literature values for similar K \cite{Rotter08,Li08} or Co \cite{Ni08,Chu09,Lester09,Pratt09,Christianson09}
contents. The diamagnetic susceptibility saturates at low temperatures. Using $\chi$ values at 2 K, the superconducting volume fraction is
estimated to be nearly 100 $\%$ for both Ba$_{0.6}$K$_{0.4}$Fe$_2$As$_2$ and Ba(Fe$_{1.95}$Co$_{0.05}$)$_2$As$_2$.

\begin{figure}[t!]
\onefigure[width=8.8cm]{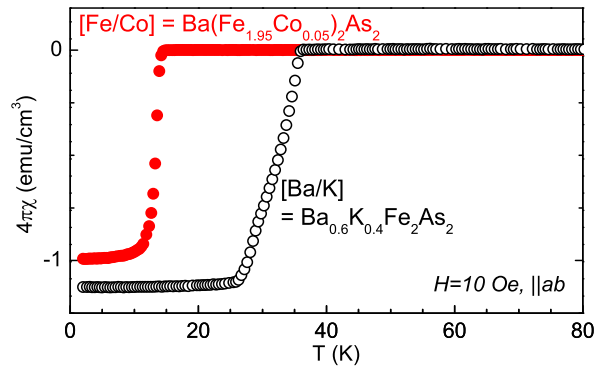} \caption{Bulk magnetization data indicating bulk superconductivity for the cobalt and potassium doped
single crystals (see text). For the [Ba/K] sample, the saturation value exceeds the theoretical value for a full Meissner volume by a
factor 1.13.}
\end{figure}

\begin{figure}[t!]
\onefigure[width=8.5cm]{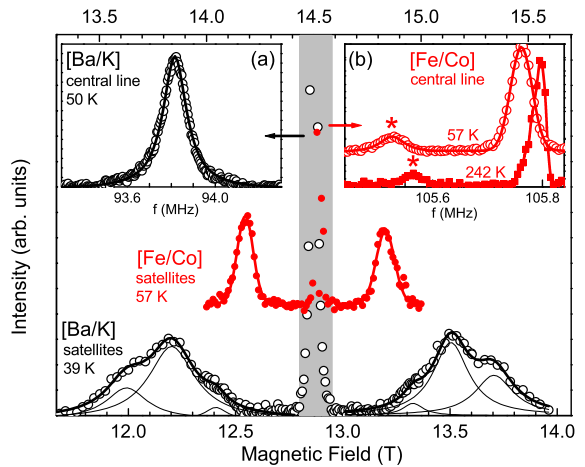} \caption{Main panel: $^{75}$As NMR spectra for $H\|c$ showing the central line (grey area,
$m_I=\frac{1}{2}\leftrightarrow -\frac{1}{2}$ transitions) and satellite lines ($m_I=\pm\frac{3}{2}\leftrightarrow\pm\frac{1}{2}$
transitions). For Ba$_{0.6}$K$_{0.4}$Fe$_2$As$_2$ (hereafter [Ba/K]), three distinct quadrupole lines are identified (continuous thin
lines are Lorentzian fits to the lineshapes). Owing to fact that the main axis of the electric field gradient tensor is the $c$-axis, the
center-to-center distance between the three pairs of Lorentzian functions defines the following quadrupole frequencies: $\nu^1_Q=6.25$~MHz
(relative area A1 = 30.5\%, full width at half maximum (FWHM) w1= 1.63~MHz), $\nu_Q(2)=4.75$~MHz (A2 = 64\%, w2 = 1.67~MHz),
$\nu_Q(3)=3.35$~MHz (A3=5.5\%, w3 = 0.85~MHz). For Ba(Fe$_{0.95}$Co$_{0.05}$)$_2$As$_2$ (hereafter [Ba/Co]), a single pair of quadrupole
satellites is observed (average FWHM w=584~kHz), the splitting of which defines $\nu_Q=2.38$~MHz (2.61 MHz at 197 K). Panels (a) and (b):
Detail of the central lines for the two samples. For [Fe/Co], the asymmetric shape of the central line at high $T$ progressively
disappears upon cooling and there is an unidentified signal accounting for $\sim$ 20\% of the total intensity (marked by the symbol
$\star$ and previously reported in~\cite{Ning08}).}
\end{figure}

\section{NMR spectra}

$^{75}$As NMR spectra are presented in Fig.~2 with quantitative information included in the caption. For both
Ba$_{0.6}$K$_{0.4}$Fe$_2$As$_2$ (hereafter abbreviated as [Ba/K]) and Ba(Fe$_{0.95}$Co$_{0.05}$)$_2$As$_2$ (hereafter [Fe/Co]), the
quadrupole parameters and the full-width-at-half-maximum (FWHM) of the NMR lines appears to be comparable to the best data published,
confirming that self-flux grown crystals are considerably cleaner than Sn-flux ones (For comparisons, see
\cite{Fukazawa09,Mukho09,Mukuda09,Matano09} and \cite{Ning08,Ning09}, respectively).

\begin{figure}[t!]
\onefigure[width=8.3cm]{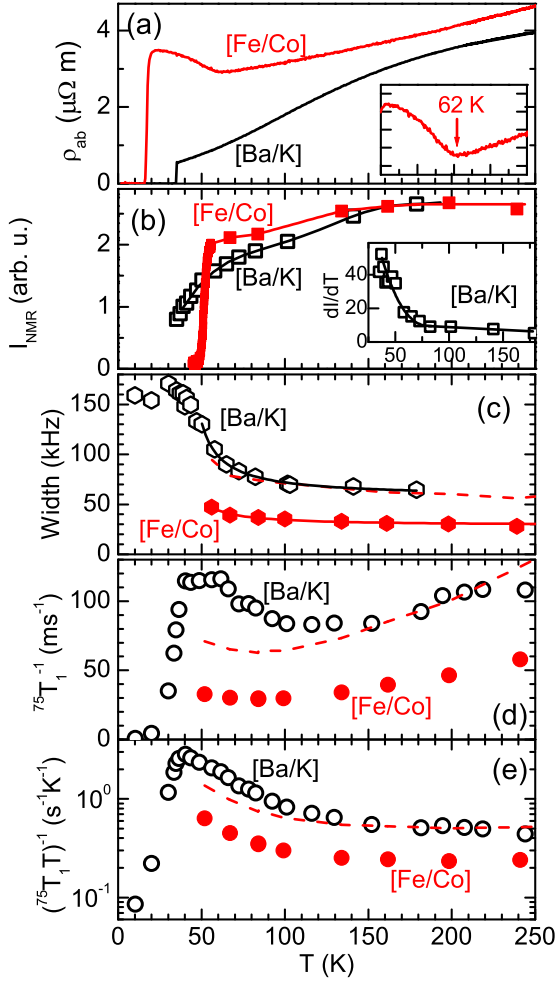} \caption{(a) In-plane resistivity of the K and Co doped samples. Inset: $\rho_{ab}$ minimum at 62~K
for the [Ba/Co] sample (SDW transition). (b) $^{75}$As NMR intensity (integrated over the full width of the central lines and normalized
by a factor $T$ accounting for the Curie dependence of the nuclear magnetization). No $T_2$ correction was necessary as $T_2$ is always
much longer than the delay separating the two pulses of the spin-echo sequence. Inset: $dI_{\rm NMR}/dT$ data for [Ba/K]. The increasing
loss of intensity below $\sim75$~K is attributed to the occurrence of magnetic order in only part of this sample. The continuous lines are
guides to the eye. (c) Width (FWHM) of the $^{75}$As NMR central lines shown in Fig.~2. The solid lines are fits to $a+b/(T+\theta)$, with
$a=57(2)$~kHz and $\theta=-38(2)$ for [Ba/K] ($T\geq 52$~K) and $a=29(2)$~kHz and $\theta=-36(2)$ for [Fe/Co]. (d) Spin-lattice relaxation
rate $T_1^{-1}$. (e) Same $T_1^{-1}$ data as in (d), divided by $T$. In panels (c), (d) and (e), the dashed lines represent the data for
[Fe/Co] multiplied by an appropriate factor in order to match the [Ba/K] data about 175~K.}
\end{figure}
%

\section{NMR intensity at high temperature}

A central result of the present work is the $T$ dependence of the total $^{75}$As NMR intensity $^{75}I_{\rm NMR}$ (Fig. 3b), which is
proportional to the number of nuclei contributing to the signal. For both samples, this intensity decreases moderately with $T$ below
$\sim200$~K. Park {\it et al.} claim the occurrence of magnetic order in $\sim$21\% (at 300~K) to 26\% (70~K) of their
Ba$_{1-x}$K$_{x}$Fe$_2$As$_2$ sample \cite{Park09}. A similar phenomenon cannot be excluded here as it could indeed cause part of the NMR
signal to disappear. However, the $T$ dependence of our NMR intensity is stronger than the $T$ variation of their magnetic volume
fraction. Furthermore, it is also likely that the penetration of the radiofrequency is limited by the skin effect in our crystals with a
few hundreds micron thickness: The relative change in $\sqrt{\rho_{ab}}$ (which is proportional to the skin depth $\delta$) is 15\% for
[Fe/Co] and 45\% for [Ba/K], between 180~K and 70~K. These numbers are comparable to the relative signal loss which amounts to 25\% and
45\%, respectively. Therefore, a significant part, if not all, of the smooth decrease of the NMR signal intensity is likely due to the
skin effect in these metallic single crystals.

Upon decreasing $T$ further, the NMR intensity for the two samples clearly displays different trends, which we discuss now.

\section{Phase coexistence in Ba(Fe$_{1.95}$Co$_{0.05}$)$_2$As$_2$}

For [Fe/Co], the signal intensity $^{75}I_{\rm NMR}$ drops abruptly to almost zero below $\sim$56~K, {\it i.e.} far above $T_c$. In the
same temperature range, $\rho_{ab}$ shows a clear upturn attributed to the appearance of a SDW phase in previous
works~\cite{Ni08,Chu09,Ahilan08}. In principle, in the SDW state, the central NMR line splits and/or broadens. So, most, if not all, of
the signal shifts away from from the paramagnetic resonance frequency $\nu_0=\gamma H_{\rm ext}$, where $H_{\rm ext}$ is the applied
external field. In this sample, despite broad magnetic field sweeps and the presence of a very small residual intensity, no well-defined
NMR signal could be measured below 50~K. This result is not surprising since extreme $^{75}$As line broadening is already visible for a Co
doping of $x=0.04$, as reported by Ning {\it et al.}~\cite{Ning09b}. Hence, the signal is "lost" and the ordered phase could not be
characterized further, except for determination of the transition temperature $T_{\rm SDW}\simeq 52\pm$3~K, defined as the point at which
the NMR signal has decreased by 50~\%, with the error bars denoting the width of the transition.

Nevertheless, the important point here is that the complete loss of NMR signal requires that every $^{75}$As nucleus has at least one
magnetic Fe site ({\it i.e.} one site in a magnetically ordered region) among its four first neighbours, within the NMR time window of
$\sim 1 \mu$s. Shifts due to longer range (dipolar) field from the more distant Fe sites can be ruled out since 1~$\mu_B$ creates a
dipolar field as low as $\sim$0.2~kOe at the next nearest As site (to be compared to the hyperfine field transferred from each first
neighbour Fe sites of $\sim$6~kOe/$\mu_B$~\cite{Grafe08}). Thus, the magnetic phase must be present {\it in all of the sample volume} at
any given time $t$. (Note that an apparent mixing due to space/time fluctuations between two macroscopically separated phases would
produce very short spin-spin relaxation time $T_2$. This is incompatible with the data at lower $x$ values~\cite{Ning09b} which suggest
that the signal is lost because of extreme broadening, not because of too short $T_2$). The presence of magnetic order over the whole
sample volume is in agreement with conclusions drawn from NMR intensity data mentioned by Ahilan {\it et al.}~\cite{Ahilan08} and from the
$\mu$SR results of Bernhard {\it et al.}~\cite{Bernhard09}. The coexistence has also been observed in neutron
scattering~\cite{Lester09,Pratt09,Christianson09}. The close values of $T_{\rm SDW}$ in neutrons and NMR indicates the absence of
probe-frequency dependence.

Since both magnetic order and superconductivity are present in the full sample volume, and NMR probes these two phases must "coexist
homogeneously", {\it i.e.} they are mixed at the microscopic (sub-nanometer) scale or even present on the same electronic sites.

\section{Phase coexistence in Ba$_{0.6}$K$_{0.4}$Fe$_2$As$_2$}

For [Ba/K], close inspection of the intensity data (Fig.3b) reveals that $^{75}I_{\rm NMR}$ decreases much more strongly below $\sim$75~K.
We attribute this additional loss of signal to the presence of magnetic order below $T_{\rm SDW}\simeq 75$~K. We note that for the same
40~\% potassium concentration, Fukazawa {\it et al.} roughly estimate 80~K~$ \leq T_{\rm SDW} \leq$~100~K from the $T$ dependence of the
NMR linewidth~\cite{Fukazawa09}, while Mukhopadhyay {\it et al.} report a loss of NMR intensity below $\sim$75~K in a single crystal
containing a significant amount of Sn impurities~\cite{Mukho09}.

The important point here is that $^{75}I_{\rm NMR}$ remains finite down to the superconducting transition at $T_c$ and even below (where
$^{75}I_{\rm NMR}$ can no longer be reliably measured). The persistence of a well defined NMR signal at the paramagnetic resonance
frequency implies that a fraction of the nuclei is not coupled to any magnetic Fe site, even on the long time scale of NMR. Hence,
magnetic order occurs {\it only in finite-size regions of the sample}. Regions without magnetic order are present at all temperatures
below 37~K and these regions clearly appear superconducting from NMR, as expected from the full Meissner fraction of the sample. Indeed, a
sharp drop of $1/T_1$ is observed below $T_c$ (Fig. 3d,e), in agreement with other results for Ba$_{1-x}$K$_x$Fe$_2$As$_2$ where $1/T_1$
was fit with one~\cite{Fukazawa09} or two~\cite{Matano09,Yashima09} gap functions. We also observed a sharp drop of the magnetic hyperfine
shift $^{75}K$ below $T_c$, indicating spin-singlet pairing (not shown).

In agreement with a number of other works~\cite{Fukazawa09,Park09,Aczel08,Rotter09,Goko09}, our results thus point to a more inhomogeneous
phase coexistence in this K-doped sample than in [Fe/Co]. The reduced magnetic fraction probably explains why no upturn is observed in the
resistivity of this sample. We point out that, since the signal does not completely disappear and no spectacular effect is seen in either
the line width or the $T_1$ data (see below), only a careful estimate of the NMR intensity is able to reveal the signal loss and thus the
presence of a magnetic state. This progressive (instead of abrupt) loss of signal at the paramagnetic resonance frequency is typical of
situations where a significant electronic inhomogeneity is present, as we shall confirm below.

\section{Spin dynamics above the magnetic transition}

For [Fe/Co], the spin-lattice relaxation rate divided by temperature $(T_1T)^{-1}$, which is proportional to a weighted sum over wave
vectors $q$ of the dynamic susceptibility $\chi^{\prime\prime}(q,\omega_{\rm NMR})$, continuously increases on cooling down to $T_{\rm
SDW}$ (Fig.~3e). From this point of view, spin dynamics above the transition appear similar to what is observed in
Ba(Fe$_{1-x}$Co$_{x}$)$_2$As$_2$ for lower $x$ values, including $x=0$~\cite{Kitagawa08,Baek08,Ning09}. Note that $^{75}$As nuclei are
located on top of the center of the square formed by four Fe sites of the same plane, so they are weakly sensitive to fluctuations
involving a transfer of wave vector $q_{x,y}=\pi/a$. This probably explains why the critical divergence in $1/T_1$, which is standard for
most antiferromagnetic transitions, is absent here. Indeed, $1/T_1$ only increases by 10~\% from 100~K to 52~K.

For [Ba/K], $(T_1T)^{-1}$ (Fig. 3e) shows a similar thermal variation as for [Fe/Co]. However, the plot of $T_1^{-1}$~$vs.$~$T$ (Fig.~3d)
reveals an important difference (besides the plateau of $1/T_1$ between 60 and 40~K): There is a 30\% increase of $1/T_1$ on cooling below
$\sim$110~K, much larger than for the Co doped sample. Interestingly, for a slightly smaller K doping value of $x=0.3$, Matano {\it et
al.} report a hump in $(T_1T)^{-1}$~$vs.$~$T$ around 90~K, which they attribute to a structural (tetragonal to orthorhombic) transition
because a small change of the quadrupole coupling is observed at the same temperature~\cite{Matano09}. For $x=0.4$, on the other hand,
Yashima {\it et al.} do not report any hump in $1/T_1$~\cite{Yashima09} and Takeshita and Kadono report the absence of frozen
magnetism~\cite{Takeshita09}. At present, given both the uncertainties on the precise doping level of the different samples and the sharp
changes in the electronic properties which probably occur close to $x=0.4$, it is difficult to comment on possible discrepancies as well
as on the presence or absence of a structural lattice distortion associated with the onset of the SDW. It is also interesting to mention a
recent study by Inosov {\it et al.}~\cite{Inosov09}: Their single crystal of Ba$_{0.6}$K$_{0.4}$Fe$_2$As$_2$, which also contains magnetic
order in finite-size regions, does not break its tetragonal symmetry macroscopically, although local distortions are clearly observed. We
speculate that remanent low energy lattice fluctuations failing to induce a long range crystallographic transition could be the origin of
the fluctuations of the hyperfine field and/or of the electric field gradients, which enhance $T_1^{-1}$ above $T_{\rm SDW}$.

\begin{figure}[t!]
\onefigure[width=8.5cm]{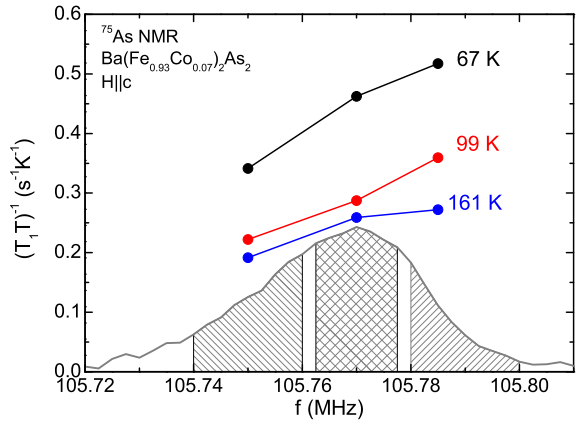} \caption{(a) Frequency dependence of $(^{75}T_1T)^{-1}$ in the [Fe/Co] sample at three selected
temperatures. Each $T_1$ value is obtained after integrating the signal over a frequency interval represented by the shaded areas on the
$^{75}$As NMR line. Note that if the signal is integrated over the whole line, a fit of the recovery data including a stretching exponent
$\alpha$ would typically yield $\alpha=0.90$ for $T=67$~K, barely distinguishable from the single-component fit $\alpha=$~1.}
\end{figure}
%

\section{Spatial inhomogeneity}

Around $\sim$200~K, the $^{75}$As NMR line width, which reflects the distribution of magnetic hyperfine shift values $^{75}K$, is much
larger for [Ba/K] than for [Fe/Co] (Fig.~3c). The distribution of quadrupole frequencies (related to the electric field gradients at the
$^{75}$As sites) is also significantly larger in the K-doped sample (Fig.~2). Evidently, there should be a correlation between the more
inhomogeneous mixing of the magnetic and superconducting phases in the K-doped sample and its larger electronic (and possibly structural)
inhomogeneity. As can be inferred from the magnetisation data in figure 1, the homogeneity of our sample is probably not optimal. However,
considering the converging recent reports~\cite{Park09,Ahilan08,Bernhard09,Aczel08,Rotter09,Goko09}, it can safely be concluded that 40\%
potassium doping creates much more inhomogeneity/disorder in the FeAs layers than $\sim$5\% cobalt doping. Upon cooling, the NMR line
broadens and follows a Curie-Weiss law for both samples (Fig.~3c). This contrasts with the small linear decrease with $T$ of the mean
value of the magnetic hyperfine shift $^{75}K$, which is proportional to the intrinsic (local) static spin susceptibility ($^{75}K$ data
for our two samples are not shown but they are identical to published data). The opposite temperature dependence of the width and the
position of the NMR line indicates that the local magnetization in FeAs layers becomes increasingly distributed at low temperature.

For [Fe/Co], information from the line width data is limited because of the abrupt disappearance of the NMR signal below $\sim52$~K.
Nevertheless, spatial magnetic inhomogeneity is also observed in the $T_1$ measurements. As shown in Fig.~4, the value of $1/T_1$ is
frequency dependent across the NMR line. Ning {\it et al.}~\cite{Ning09} obtained quantitatively similar $T_1$ data for several Co
contents and they could correlate this frequency dependence with the local Co doping level. In other words, there is a significant spatial
distribution of the electron density in the FeAs layers, which is also inferred from STM~\cite{Massee09}. Other sources of line
broadening, such as a (paramagnetic) staggered magnetisation, may be present as well.

For [Ba/K], a similar $T_1$ analysis was not performed because the line was much broader than the excitation width provided by the NMR
pulses. Still, $T_1$ inhomogeneity is also present as a phenomenological stretching exponent $\alpha$, introduced in the fit of the
recovery data~\cite{Mitrovic08}, deviates from $\alpha=1$ below $\sim$50~K (not shown). It should be pointed out that our NMR data above
$T_{\rm SDW}$ in the [Ba/K] sample do not show the coexistence of two distinct signals either in the spectra or in the recovery laws. So,
they do not provide any argument in favor of a separation into two macroscopically distinct phases. In particular, in contrast with {\it
e.g.} CaFe$_{1-x}$Co$_x$AsF~\cite{Takeshita09}, the magnetic phase here does not appear to originate from totally undoped regions.

The most interesting result for this sample, however, is the fact that the line width remains almost constant below the superconducting
transition. The absence of narrowing below $T_{\rm SDW}$, and especially below $T_c$, suggests that the main source of broadening is
present in the non-magnetic regions of the sample. A distribution of doping values appears unlikely since the NMR shift drops to zero as
$T\rightarrow0$ in the superconducting state. A plausible scenario is that the line is broadened by a (paramagnetic) staggered spin
polarization, such as observed in Zn doped YBa$_2$Cu$_3$O$_7$~\cite{Julien00,Ouazi06,Alloul09} or in LSCO~\cite{Julien99}.

\section{Conclusion and Discussion}

To sum up our results, Ba(Fe$_{1.95}$Co$_{0.05}$)$_2$As$_2$ and Ba$_{0.6}$K$_{0.4}$Fe$_2$As$_2$ both show the coexistence of bulk
superconductivity with magnetic (SDW) order. Yet, the details appear different:

In Ba(Fe$_{1.95}$Co$_{0.05}$)$_2$As$_2$, the two phases coexist at the microscopic scale probed by NMR, a situation which is often
referred to as "homogeneous mixing" in the literature. This implies either that both types of orders are simultaneously defined at each Fe
site (owing to the multiple bands present at the Fermi level), or that they are mixed on the scale of one or two lattice spacing. A
nanoscale coexistence involving superconducting islands (without magnetic order) of typical size defined by the coherence length
$\xi\simeq2.8$~nm~\cite{Yin09} appears to be unlikely. In this case, some paramagnetic NMR signal from regions as large as ten times the
Fe-Fe distance should be observed.

In Ba$_{0.6}$K$_{0.4}$Fe$_2$As$_2$, on the other hand, the magnetic regions do not occupy the full sample volume. In this sense, the
coexistence might be qualified as "inhomogeneous".

Both samples show some spatial inhomogeneity of their magnetic properties, already above the spin-density wave transition. It is not yet
clear to what extent the inhomogeneous magnetisation in both systems is due to spatial variations of the electronic density or only of the
spin polarization. The latter may itself be a precursor manifestation of the spin-density-wave state or it may be caused by the magnetic
screening of impurities/defects~\cite{Mukho09,Julien00,Ouazi06,Alloul09}. The dramatic effect of a small amount of impurities on the
NMR-line broadening already in the undoped material~\cite{Kitagawa08,Baek08} is a strong indication that the extreme NMR broadening
observed here and by Ning {\it et al.}~\cite{Ning09b} below the transition at $T_{\rm SDW}$ is due fact that the spin-density wave state
in superconducting samples is highly disordered/inhomogeneous.

Perhaps unexpectedly, the inhomogeneity is considerably stronger in the potassium-doped sample. Actually, the fact that substitutions at
the Fe site, unlike substitutions at the Cu site in the cuprates, improves conductivity and even induces superconductivity is one of the
most remarkable surprises of these new superconductors. It is impossible to fully understand the Ba$_{1-x}$K$_{x}$Fe$_2$As$_2$ system from
the present NMR data only. The important inhomogeneity/disorder could be due to inhomogeneity of $K^+$ concentration and/or to a
particularly strong impact of these ions on the local electronic structure in FeAs layers. No evidence for phase separation could be
detected in the K-doped sample, and the single crystal exhibits 100\% Meissner fraction. Our NMR data thus seem to be in better agreement
with the picture proposed by Rotter {\it et al.}~\cite{Rotter09} than with the pure phase separation picture of Park {\it et
al.}~\cite{Park09}. We point out that $T_{\rm SDW}$ {\it vs.} $x$ is extremely steep near $x=0.4$. Phase separation or inhomogeneous
coexistence could thus originate from K-doping inhomogeneity around this particular concentration. They might thus not reflect the
properties at somewhat lower $x$ values. Said differently, Ba$_{1-x}$K$_{x}$Fe$_2$As$_2$ should perhaps be simply viewed as a disordered
version of Ba(Fe$_{1-x}$Co$_{x}$)$_2$As$_2$.

Because the magnetic transition temperature drops continuously to zero about the middle of the superconducting dome for
Ba$_{1-x}$K$_{x}$Fe$_2$As$_2$ and Ba(Fe$_{1-x}$Co$_{x}$)$_2$As$_2$, the phase diagram of these pnictides shows a clear analogy with some
heavy fermion compounds (and possibly electron-doped cuprates~\cite{Armitage09}). The coexistence of magnetic order and superconductivity
has been observed with NMR in several heavy fermions~\cite{Kitaoka02,Curro09}, the two phases being homogeneously mixed in some of them,
separated in others. There is also some analogy between the present results (see also~\cite{Nakai08,Julien08}) and the coexistence between
magnetic order and superconductivity as seen by NMR in hole-doped cuprates~(see \cite{Mitrovic08,Julien03} and references therein). On the
other hand, the data reported here do not reveal the phase separation observed with NMR in 2D~\cite{Lefebvre00} and 1D~\cite{Lee05}
organic conductors. This correlates with the fact that these materials have a quite different phase diagram.

Clearly, more experimental and theoretical work is needed to clarify all aspects of the mixing of magnetic and superconducting phases in
the pnictides. Nevertheless, their coexistence over a substantial portion of the phase diagram clearly emerges as a cornerstone of their
physics.

{\it Note added}: After the completion of this paper, we became aware of a recent NMR work [Y. Laplace {\it et al}, arXiv:0906.2125], with
similar conclusions regarding the mixing of superconducting and spin density wave orders at the microscopic scale for the cobalt-doped
system. In their more strongly (6\%) Co doped single crystal ($T_{\rm SDW}$=31~K, $T_c$=22~K), the average moment is strongly reduced and
the NMR signal in the magnetic phase is severely broadened but remains observable, in contrast with our case.

\acknowledgments

This work was supported by the Ministry of Science and Technology of China, and Chinese Academy of Sciences.

The authors wish to thank Andr\'e Sulpice (Institut N\'eel, Grenoble) for performing a squid measurement.

\end{document}